\documentstyle[]{article}

\begin{document}
\begin{center} {\Large {\bf States insensitive to the
Unruh effect in multi-level detectors
}}\\[1cm]
Karl-Peter Marzlin
\footnote{Address after June 1st, 1997:
	School of Mathematics, Physics,
	Computing and Electronics, Macquarie University, NSW 2109,
	Australia.}
and J\"urgen Audretsch
\footnote{e-mail: juergen.audretsch@uni-konstanz.de}
\\[2mm]
Fakult\"at f\"ur Physik
der Universit\"at Konstanz\\
Postfach 5560 M 674\\
D-78434 Konstanz, Germany
\end{center}
$ $\\[3mm]
\begin{minipage}{15cm}
\begin{abstract}
We give a general treatment of
the spontaneous excitation rates and the non-relativistic Lamb
shift of constantly accelerated multi-level atoms as a model for
multi-level detectors.
Using a covariant formulation of the dipole coupling between the atom and
the electromagnetic field we show that new Raman-like transitions
can be induced by the acceleration.
Under certain conditions these transitions
can lead to stable ground and excited states which are not affected by the
non inertial motion. The magnitude of the Unruh effect is not altered by
multi-level effects. Both the spontaneous excitation rates and the Lamb
shift are not within the range of measurability.
\end{abstract}
\end{minipage}
$ $ \\[1cm]
%\newpage
%%%%%%%%%%%%%%%%%%%%%%%%%%%%%%%%%%%%%%%%%%%%%%%%%%%%%%%%%%%%%%%%
\section{Introduction}
The Unruh effect \cite{unruh76} is the prediction that
a linearly accelerated two-level particle detector which is sensitive to
detect real massless perticles becomes excited when moving through the
Minkowki vacuum. It behaves as
if it were in a thermal bath of particles with
Unruh temperature $T_U = \hbar a /(2\pi k_B c)$, where $a$ is the acceleration
of the detector. Although this temperature is in general extremely small
the Unruh effect gained much interest in theoretical physics.
For a detailed discussion see for example Ref. \cite{takagi86}. After
Unruh and Wald \cite{unruh84} reported acausal correlations and lack
of energy conservation the emission of radiation by the detector has
been questioned \cite{gegenunruh}. The discussion was resolved by the
inclusion of the measurement process \cite{mueller94a}. For a two-level
atom coupled to a scalar field the
spontaneous excitation of a uniformly accelerated atom in its ground state
was interpreted in terms
of vacuum fluctuations and radiation reaction by Audretsch and M\"uller
\cite{mueller94b}.
The associated Lamb shift was worked out in Ref. \cite{aumue1},
and arbitrary stationary trajectories were discussed in Ref.
\cite{holzmann95}.
This quantum optical approach is similar to the one used in the present
paper. The experimental realizability  was discussed by Bell and Leinaas
\cite{bell87} who considered the spin of an electron in a storage ring
instead of a detector model. Since under these conditions the acceleration
can be extremely large the Unruh temperature should reach values of about
1000 K, but unfortunately the fluctuations of the electron's trajectory
in the storage ring cover the effect.

It remains therefore a challenge to propose quantum systems for which the
Unruh effect or a similar effect becomes measurable. In this connection
it has been conjectured that multi-level atoms with their variety of complex
transitions may show acceleration-induced effects which are considerably
bigger than the usual
Unruh effect. In this paper we will give an answer to this
question. To discuss realistic situations as far as the coupling is
concerned, we base our calculations on the relativistic generalization
of the electromagnetic dipole coupling. This will give as a by-product
an improved expression for the Lamb shift of accelerated atoms.

The paper is organized as follows. In Sec. 2 we introduce the dipole
coupling to the electromagnetic field for arbitrary atom trajectories. In
Sec. 3 we derive the Master equation for the density matrix of moving atoms.
For later use we review the Unruh effect for the electromagnetic coupling
in Sec. 4 and derive the corresponding Lamb shift. Finally we turn to
acceleration-induced multi-level effects in Sec. 5.
%%%%%%%%%%%%%%%%%%%%%%%%%%%%%%%%%%%%%%%%%%%%%%%%%%%%%%%%%%%%%%%%
\section{The coupling}
We consider atoms forced to move on a given classical trajectory $z^\mu (t
)$ with proper time $t$. The interaction with the electromagnetic
field is described in the dipole approximation and
consists of a relativistic generalization of the dipole coupling
\cite{borde83,takagi86},
\begin{equation}
H_{int} = - c \dot{z}^\mu(t) d^\nu F_{\mu \nu}(z^\nu(t))\; .
\label{coupling} \end{equation}
$\mu, \nu, \ldots = 0,\ldots ,3$. We use the summation convention
throughout the paper if not otherwise stated.
A dot denotes the derivative
with respect to $t$. $F_{\mu \nu}$ is the
electromagnetic field strength operator and $d^\nu$ is the dipole operator
which describes the internal induced dipole moments of the atom.

If one replaces $F_{\mu \nu}$ by the electric and magnetic field one can see that, in the laboratory frame, the coupling
(\ref{coupling}) does not only include the usual dipole coupling to the
electric field but also the R\"ontgen term \cite{roentgen} which represents
a coupling between the magnetic field, the atomic center-of-mass momentum,
and its internal  dipole moment. Its importance for the relativistic
invariance of the spontaneous decay rate has been demonstrated by Wilkens
\cite{wilkens94}. A microphysical derivation of the dipole coupling and
the R\"ontgen term including the effect of a weak gravitational field was
done in Ref. \cite{marzlin95}.

To describe the 4-velocity and the local rest frame of the atom, we introduce
along the trajectory a set of four orthonormalized vectors $e^\mu_{
\underline{\alpha}}(t)$, $\underline{\alpha}=0,1,2,3,$ with
$e_{\underline{\alpha} \mu} e_{\underline{\beta}}^\mu =$ diag$(-1,1,1,1)$
(see, e.g., Ref. \cite{misner73}). Underlined indices represent tetrad
numbers.
$e_{\underline{0}}^\mu (t)$ is the tangent $\dot{z}^\mu(t)$ to the
worldline. The three spacelike vectors
$e_{\underline{a}}^\mu(t), \; \underline{a}=1,2,3$,
are Fermi propagated along the trajectory.
With respect to the tetrad the dipole operator is given by
$d_{\underline{\alpha}} = e_{\underline{\alpha}}\cdot d \equiv
e_{\underline{\alpha}}^\mu d_{\mu}$. In the Schr\"odinger picture the
internal dipole moments are constant.
We go to the interaction picture.
The Schr\"odinger equation then reads
\begin{equation}
i \hbar \frac{d\psi}{dt} = - c d^{\underline{a}}(t)\,
	F_{\underline{0}\underline{a}}(z^\mu(t))\psi  \; .
\label{schroed} \end{equation}
Because $F_{\underline{0}\underline{0}}$ always vanishes the expression
$d^{\underline{\alpha}}(t)\,  F_{\underline{0}\underline{\alpha}}$
could be reduced to  $d^{\underline{a}}(t)\,  F_{\underline{0}\underline{a}}$.
Equation (\ref{schroed}) is manifestly covariant as well as gauge invariant.

The components $d^{\underline{a}}(t)$ of dipole moment operator
can be written as
\begin{equation}
d^{\underline{a}}(t) = \sum_{i,j=1}^N d^{\underline{a}}_{
ij} | i \rangle \langle j | \exp [i \omega_{ij} t]\; ,
\end{equation}
with
$|i \rangle \; , \; i,j =1, \ldots , N$ are the time independent internal
energy eigenstates of an atom with N internal levels with energy
$E_i$. We assume that the effect of the acceleration on the internal
states and the energy levels has already been included.
The transition frequencies are given by $\omega_{ij} :=
(E_i -E_j)/\hbar$.
Since we are working in the interaction picture
each coherence operator is oscillating
with frequency $\omega_{ij}$, and since the Schr\"odinger
equation (\ref{schroed}) is written with respect to the atom's proper
time these oscillations have still the form $ \exp [i \omega_{ij}
t]$.

The vector components  $d^{\underline{a}}_{ij}$ are constant
because they refer to the frame $e^\mu_{\underline{\alpha}}  $ attached to
the atom. They are given by
\begin{equation}
\vec{d}_{ij}= e \langle i | \vec{x} | j \rangle
\label{matelem} \end{equation}
for a one-electron atom and where $e$ is the charge of the electron and
$\vec{x}$ its position operator relative to the center-of-mass position.
Any spatial vector $\vec{v}$ is defined by having the components
$v^{\underline{b}}\; ,\; \underline{b}=1,2,3, $ with respect to the tetrad.
Because $d^{\underline{a}}(t) $ has to be hermitean we have
$d^{\underline{a}}_{ij} = d^{\underline{a}*}_{ji}$ and
because of parity conservation $d^{\underline{a}}_{ii} =0$ holds.

To quantize the Maxwell field we introduce the electromagnetic four-potential
$A_\mu$ with $F_{\mu \nu} =\partial_{\mu} A_{\nu} - \partial_{\nu} A_{\mu}$.
The laboratory frame is an inertial reference frame. With regard to the
concept of a photon and accordingly to the concept of the {\em
Minkowski vacuum},
we refer to this frame and decompose the four-potential in the Lorentz gauge
as follows:
\begin{equation}
	A^\mu (x^\nu) = \sqrt{\frac{\hbar}{2 (2\pi)^3 \varepsilon_0 c}}
	\int {d^3k \over \sqrt{k}} \Big \{ \exp (i k \cdot x)
	\varepsilon_{\hat{\alpha}}^\mu ({\bf k})
	a^{\hat{\alpha}}({\bf k}) + H.c. \Big \}\; ,
\label{amu}\end{equation}
with $k^\mu := (k\equiv |{\bf k}|,{\bf k})$ and ${\bf k}$ being the
wavevector in the laboratory frame.
$a^{\hat{\alpha}}({\bf k})$ are the usual annihilation
operators fulfilling $ [a^{\hat{\alpha}}({\bf k}),
a^{\hat{\beta}}({\bf k}^\prime)] = \delta ({\bf k}-
{\bf k}^\prime) \eta^{\hat{\alpha} \hat{\beta}}$.
The vectors $\varepsilon_{\hat{\alpha}}^\mu ({\bf k})$ are
the (real) normalized polarization vectors of the electromagnetic field
and hatted indices are polarization indices,
$\hat{\alpha} = 0,1,2,3$.
$\varepsilon_{\hat{0}}^\mu ({\bf k})$ corresponds to the
scalar photon, $\varepsilon_{\hat{3}}^\mu ({\bf k})$ to the
longitudinal photon, and $\varepsilon_{\hat{1}}^\mu ({\bf k})$
as well as $\varepsilon_{\hat{2}}^\mu ({\bf k})$ to the two
transverse degrees of freedom related to the physical photons
(compare, e.g., Ref. \cite{itzykson80}).
Note that for each ${\bf k}$ these four vectors do provide an orthonormal
tetrad at rest adjusted to the photon ray with wavevector ${\bf k}$. When
working out the influence of the acceleration of the atoms we will
not only have to take into account
their non-inertial four-trajectory but also
the orientation of the rest-frame tetrad
$e_{\underline{\alpha}}^\mu(t)$ relative to the laboratory tetrad
$\varepsilon_{\hat{\alpha}}^\mu ({\bf k})$. On this basis we will now
formulate the quantum optics of multi-level atoms moving with uniform
acceleration through the Minkowski vacuum $|0 \rangle$ with
$a^{\hat{\alpha}}({\bf k}) |0 \rangle = 0$.
%%%%%%%%%%%%%%%%%%%%%%%%%%%%%%%%%%%%%%%%%%%%%%%%%%%%%%%%%%%%%%%%
\section{Master equation for moving atoms}
In this paper we
will use a master equation to calculate the decay rates and the
non-relativistic part of the Lamb shift for a given energy level. The
master equation technique is described in many textbooks on
quantum optics (see, e.g., Ref. \cite{cohen92}). In short its
central assumption is that the density matrix $\rho_{tot}$ for the
atom-field system can approximately be decomposed into an
atomic density matrix $\rho(t)$ and a field density matrix
(in our case the Minkowski vacuum) which
remains constant, $ \rho_{tot} (t) = \rho (t) \otimes | 0 \rangle \langle
0|$. In second order of perturbation theory
one can derive the following master
equation for the atomic density matrix,
\begin{eqnarray}
\frac{d \rho (t)}{d t} &=&\frac{-1}{\hbar^2}
\int_0^t dt^\prime
\Big \{ [ d^{\underline{a}}(t), d^{\underline{b}}(
t^\prime)\, \rho(t)] \, G_{\underline{a} \underline{b}}
(t,t^\prime) \nonumber \\ & & \hspace{1.5cm} -
[ d^{\underline{a}}(t), \rho(t)\, d^{\underline{b}}(
t^\prime) ] \, G_{\underline{a} \underline{b}}^*
(t,t^\prime) \Big \}\; ,
\label{master} \end{eqnarray}
with the correlation function
\begin{equation}
G_{\underline{a}\underline{b}}(t,t^\prime) := c^2
\langle 0 | F_{\underline{0}\underline{a}}(z^\nu(t)) \;
F_{\underline{0}\underline{b}}(z^\nu(t^\prime)) | 0 \rangle \; .
\label{tpij} \end{equation}
Using Eqs. (\ref{amu}) the function $
G_{\underline{a}\underline{b}}(t,t^\prime)$ can be
written as
\begin{eqnarray}
G_{\underline{a}\underline{b}}(t,t^\prime)&=& \frac{\hbar c}{2
  (2\pi)^3 \varepsilon_0} \int {d^3k \over k} e^{i k\cdot (z (t)-
  z (t^\prime))} \Big \{ k_{\underline{0}}(t) k_{\underline{0}}(t^\prime)
  (e_{\underline{a}}(t) \cdot e_{\underline{b}}(t^\prime))
  - \label{galbet} \\ & &
  k_{\underline{0}}(t) k_{\underline{b}}(t^\prime)
  (e_{\underline{a}}(t) \cdot e_{\underline{0}}(t^\prime)) -
  k_{\underline{a}}(t) k_{\underline{0}}(t^\prime)
  (e_{\underline{0}}(t) \cdot e_{\underline{b}}(t^\prime)) +
  k_{\underline{a}}(t) k_{\underline{b}}(t^\prime)
  (e_{\underline{0}}(t)\cdot e_{\underline{0}}(t^\prime))
\Big \}\, .
 \nonumber\end{eqnarray}
Its precise form depends on the atomic trajectory $z^\mu(t)$.
The trajectory manifests itself also in the projections on the comoving
(Fermi propagated) tetrad.
To solve the master equation it is convenient to expand $\rho$
as
\begin{equation}
\rho (t) = \sum_{i,j=1}^N |i \rangle \langle j| \rho_{ij}(t)
\end{equation}
and to insert this in Eq. (\ref{master}). We will describe the stationary
situation of inertially moving atoms or atoms in constant acceleration. The
time dependence in Eq. (\ref{galbet}) reduces therefore to $t^{\prime
\prime} = t-t^\prime$ and we can make
the commonly used approximation to extend the resulting integral
from $0$ to $\infty$ instead from $0$ to $t$. This is well
justified if $G_{\underline{a}\underline{b}}(t)$
falls off rapidly. Defining
\begin{equation}
K_{ij,kl}(\omega) :=
  d^{\underline{a}}_{kl}d^{\underline{b}}_{ij}
  \int_0^\infty dt^{\prime \prime} e^{-i \omega t^{\prime
  \prime}} G_{\underline{a}\underline{b}}(t^{\prime
  \prime})
\label{kijkl} \end{equation}
one finds the master equation
\begin{eqnarray}
  \frac{d \rho_{ij}}{d t} &=& \frac{-1}{\hbar^2} \sum_{k,l=1}^N
  \Big \{ \rho_{lj} e^{i \omega_{il}t} K_{ik,kl} (\omega_{kl})+
  \rho_{ik} e^{i \omega_{kj}t} K_{jl,lk}^* (-\omega_{kl})
  \nonumber \\ & & \hspace{1.3cm}
  - \rho_{kl} e^{i t(\omega_{ik}+ \omega_{lj})} \big [
  K_{lj,ik} (\omega_{ik}) + K_{ki,jl}^* (-\omega_{lj}) \big ]
  \Big \} \; .
\label{rhomaster}\end{eqnarray}
We will see below that the two-point function (\ref{tpij}) is isotropic,
$G_{\underline{a}\underline{b}}(t)=\delta_{\underline{a}\underline{b}} G(t)$.
Accordingly we can reduce Eq. (\ref{kijkl}) to
\begin{equation}
K_{ij,kl}(\omega) = (\vec{d}_{ij}\cdot \vec{d}_{kl})\, K(\omega) \; .
\label{isotrop} \end{equation}
The specific trajectory will manifest itself in the form of the function
$K(\omega)$.
%%%%%%%%%%%%%%%%%%%%%%%%%%%%%%%%%%%%%%%%%%%%%%%%%%%%%%%%%%%%%%%%
\section{Unruh effect and Lamb shift for two-level atoms}
We study two-level atoms first to derive the modification of the Lamb shift
caused by the acceleration for the interaction of the atom with the
inertial electromagnetic instead of the scalar vacuum.
In doing so we will as well provide
the expressions on which the discussion of multi-level atoms can be based.
We briefly
review the inertial situation as well because we have to refer to it for
the renormalization.

For a two-level atom with an excited state $|e \rangle$ and a ground state
$| g \rangle$ two
components in the dipole operator, namely $d^{\underline{a}}_{eg}$
and $d^{\underline{a}}_{ge} = d^{\underline{a}*}_{eg}$,
do not vanish. The master equation (\ref{rhomaster}) then becomes
\begin{eqnarray}
\frac{d \rho_{gg}}{dt} &=& \frac{2|\vec{d}_{eg}|^2}{\hbar^2}
  \Big \{ \rho_{ee} \mbox{ Re } K(-\omega_{eg}) - \rho_{gg}
  \mbox{ Re } K(\omega_{eg}) \Big \}
\label{rhogg} \\
\frac{d \rho_{eg}}{dt} &=& -\rho_{eg} \frac{|\vec{d}_{eg}|^2}{\hbar^2}
  \big [ K(-\omega_{eg}) + K^*(\omega_{eg})\big ]
  +\rho_{ge} \frac{\vec{d}_{eg}^2}{\hbar^2} e^{2i \omega_{eg}t}
  \big [ K(\omega_{eg}) + K^*(-\omega_{eg})\big ]
\label{rhoeg} \end{eqnarray}
together with the normalization condition
$\rho_{ee}+ \rho_{gg} =1$. The term proportional
to $\rho_{ee}$ in Eq. (\ref{rhogg}) describes the spontaneous
emission of a photon and the corresponding deexcitation of the atom
and will be present also for inertially moving atoms. The term
proportional to $\rho_{gg}$ describes spontaneous excitation.
It vanishes for inertial atoms and displays the Unruh effect for
accelerated atoms. Eq. (\ref{rhoeg}) is mainly of interest for
the imaginary part of the coefficent connected with $\rho_{eg}$.
In the Schr\"odinger picture the unperturbed coherence $\rho_{eg}$
oscillates with the transition frequency $\omega_{eg}$ and hence
gives us information about the energy difference between the two
states. An imaginary part in the interaction picture indicates
that the oscillation frequency and therefore the energy difference
has changed. It describes the level shift caused by the interaction
with the atom and can be interpreted as the non-relativistic
Lamb shift.

{\em Inertial atoms:}
The worldline of the inertially moving atom is
given by $z^\mu (t) = \bar{z}^\mu + c t e_{\underline{0}}^\mu$.
The tetrad vectors $e_{\underline{\alpha}}^\mu$ are constant for this
trajectory.
Changing in Eq. (\ref{galbet}) the integration variables to $p_c :=
k_{\underline{c}}$ one can
exploit the invariance of the integration measure under
Lorentz boosts \cite{jacobiproof}, i.e., $d^3 k/k = d^3 p/p$ with $p :=
|\vec{p}| = -k_{\underline{0}}$, to simplify Eq. (\ref{tpij})
to $G_{\underline{a}\underline{b}}^{\mbox{\scriptsize inertial}} =
\delta_{\underline{a}\underline{b}} G^{\mbox{\scriptsize inertial}}$
with
\begin{equation}
G^{\mbox{\scriptsize inertial}}
  (t-t^\prime) = \frac{-\hbar}{6 \pi^2 \varepsilon_0 c^3}
  \frac{d^3}{dt^3} \Big \{
  i\pi \delta (t-t^\prime) + \frac{{\cal P}}{t-t^\prime} \Big \} \;.
\end{equation}
Here ${\cal P}$ denotes the principal value (in the sense of
distributions) and we have used
\begin{equation}
\int_0^\infty dp e^{-i p x} = \pi \delta (x) -i \frac{{\cal P}}{x}
\; .
\label{formel} \end{equation}
Inserting this into Eq. (\ref{isotrop}) and using
$G_{\underline{a}\underline{b}}^*(t) =
G_{\underline{a}\underline{b}}(-t)$ (for all cases
considered in this paper) we arrive at
\begin{equation}
2 \mbox{ Re } K^{\mbox{\scriptsize inertial}}(\omega) =
   \int_{-\infty}^\infty dt e^{-i \omega t}
   G^{\mbox{\scriptsize inertial}}(t) = \theta (-\omega) \frac{\hbar
   |\omega|^3}{3 \pi \varepsilon_0 c^3} \quad ,\; i,j =1,2,3.
\label{ink2} \end{equation}
In the derivation we used the residue theorem and
the fact that for functions $f(z)$
with a first order pole at $z=0$ the expression $\int_{-\infty}
^\infty dx {\cal P} f(x)$ can be replaced by
$(1/2)\lim_{\varepsilon\rightarrow 0}\int_{-\infty}
^\infty dx \{ f(x+i \varepsilon) + f(x-i \varepsilon)\}$.
According to Eq. (\ref{rhogg}) the spontaneous emission rate of an excited
atom is given by the first term in Eq. (\ref{rhogg}),
$ 2|\vec{d}_{eg}|^2$ Re $K(-\omega_{eg})/ \hbar^2$, which results in
\begin{equation}
\Gamma_{\mbox{\scriptsize inertial}}^{e\rightarrow g} =
   \frac{\omega_{eg}^3
   |\vec{d}_{eg}|^2}{3\pi \hbar \varepsilon_0 c^3} \; .
\label{ink} \end{equation}
This coincides with the classic result of Wigner and Weisskopf
(see, e.g., \cite{milonni95}).
On the other hand, we can infer from the second term in Eq. (\ref{rhogg})
that inertial atoms in vacuum do not suffer from spontaneous excitation
in the ground state
because the corresponding transition rate $\Gamma_{\mbox{\scriptsize inertial}}^{g\rightarrow e}$ is proportional to Re $K^{\mbox{\scriptsize inertial}}(\omega_{eg})$ and thus vanishes because of $\theta (-
\omega)$ in Eq. (\ref{ink2})

We have to remark that the imaginary part of
$K^{\mbox{\scriptsize inertial}}(\omega)$,
\begin{equation}
  \mbox{Im }K^{\mbox{\scriptsize inertial}}(\omega) = \frac{\hbar \omega^3}{
  6\pi^2 \varepsilon_0 c^3} \int_0^\infty dt \frac{{\cal  P}}{t} \cos
  (\omega t) \; ,
\end{equation}
is divergent. To renormalize it we refer to the previously mentioned fact that
the imaginary part corresponds to an energy level shift
$\Delta E_{\mbox{\scriptsize inertial}}$. Therefore, if we set
as {\em renormalization} Im $K^{\mbox{\scriptsize
inertial}}(\omega) =0 $, the internal energies $E_e$ and $E_g$ are not the
bare energy eigenvalues but include the radiative level shift so that
$\Delta E_{\mbox{\scriptsize inertial}}=0$. We will see
that for an accelerated atom this radiative energy shift will be changed.
%%%%%%%%%%%%%%%%%%%%%%%%%%%%%%%%%%%%%%%%%%%%%%%%%%%%%%%%%%%%%%%%

{\em Accelerated atoms:}
We now turn to uniformly accelerated atoms moving
through the Minkowski vacuum on the well known Rindler trajectory
\begin{equation}
z^\mu(t) = \frac{c^2}{a} (\sinh [at/c],0,0,\cosh [at/c])\; ,
\end{equation}
where $a>0$ is the constant
acceleration. The tetrad vectors $e_{\underline{1}}^\mu$
and $e_{\underline{2}}^\mu$ carried with the atom do always point in the
1- and 2-direction of the laboratory frame. The remaining vectors are given by
$ e_{\underline{0}}^\mu = (\cosh [at/c] ,0,0, \sinh [at/c])$ and
$ e_{\underline{3}}^\mu = (\sinh [at/c] ,0,0,\cosh [at/c])$.
The two-point function can be calculated similarly to the inertial case
by changing the integration variables in Eq. (\ref{galbet}) to $p_i
= k_{\underline{i}}((t+t^\prime)/2)$.
The two-point functions  have been derived in section 9.3
of Ref. \cite{takagi86} in the context of a detector model with the result
\begin{equation}
c^2 \langle 0|F_{\underline{0}\underline{a}}(t)
         F_{\underline{0}\underline{b}}(t^\prime) |0\rangle = \delta_{
	 \underline{a}\underline{b}} G(t-t^\prime)\; ,
\label{2punkts}\end{equation}
where within our model the distribution $G$ is defined by
\begin{equation}
G(t-t^\prime) := \frac{-\hbar a^4}{96\pi^2\varepsilon_0 c^7}
  \frac{d^3}{du^3} \left .\left \{
  i \pi \delta (u)+ \frac{{\cal P}}{u} \right \} \right |_{u =
  \sinh (\Delta)},
\label{gijende}\end{equation}
with $\Delta := (t-t^\prime)a/(2c)$.
The derivative of the $\delta$ function and the principal value is defined
in the sense of distributions: Let $D(x)$ be a distribution, $g(x)$ a
test function, and $f(x)$ some function which is monotonic in the interval
of integration. Then the derivative is defined by
\begin{equation}
\int dx g(x) \frac{d^n}{du^n} D(u) \big |_{u=f(x)} = (-1)^n \int dx
  D(f(x)) \left ( \frac{d}{dx} \frac{1}{df/dx} \right )^n g(x)\; .
\label{distrib} \end{equation}

Takagi \cite{takagi86} also derived the power spectrum 2 Re
$K (\omega)$ for the detector model. Transformed
into our notation it reads
\begin{eqnarray}
2 \mbox{ Re } K(\omega) &=&
  \frac{\hbar |\omega|^3}{3\pi \varepsilon_0 c^3}
  \left [ 1 + \frac{a^2}{c^2 \omega^2}\right ]
  \Big \{ \theta(-\omega) + \frac{1}{e^{2\pi |\omega| c/a}-1}\Big \}
\label{rek}
\end{eqnarray}
The spontaneous transition rates can be deduced by inserting this into
Eq. (\ref{rhogg}) and are given by
\begin{eqnarray}
\Gamma^{e\rightarrow g}_{\mbox{\scriptsize local}} &=& \Gamma_{\mbox{\scriptsize inertial}}
   \left [ 1 + \frac{a^2}{c^2 \omega_{eg}^2}\right ]
  \Big \{ 1 + \frac{1}{e^{2\pi \omega_{eg} c/a}-1}\Big \}
  \label{emrate} \\
\Gamma^{g\rightarrow e}_{\mbox{\scriptsize local}} &=& \Gamma_{\mbox{\scriptsize inertial}}
   \left [ 1 + \frac{a^2}{c^2 \omega_{eg}^2}\right ]
  \frac{1}{e^{2\pi \omega_{eg} c/a}-1}\; .
\label{absrate} \end{eqnarray}
The "1" in the curly brackets of Eq. (\ref{emrate}) is just the inertial
decay rate while the second term in these brackets represents the
thermal factor used to define the Unruh temperature $T_U$ by
$(\hbar |\omega|)
/(k_B T_U) = 2\pi |\omega| c/a$. The additional term in square brackets
is a non-thermal correction. It is
caused by the coupling to the electromagnetic field strength operator
instead of a structureless scalar field. The possibility of a
spontaneous excitation ($\Gamma^{g\rightarrow e}_{\mbox{\scriptsize local}}
> 0$) in the ground state
corresponds to that phenomenon which is usually referred to as
the {\em Unruh effect}. The non thermal corrections
for the electromagnetic dipole coupling are not in conflict with the proof
that the Unruh effect is of thermal
nature for all quantum fields \cite{bell85} because this was shown
for the equilibrium partition only. The precise statement of the proof is
that, for a
two-level atom for instance, the equilibrium partition obeys $\rho_{ee}/
\rho_{gg} = \exp [- \hbar  \omega_{eg}/(k_B T_U)]$. It is easy to see that
this relation is fulfilled by the stationary solution of Eq. (\ref{rhogg}).

Another physically interesting effect is the radiative level shift
of the accelerated atom for the coupling to the Maxwell field.
For a two-level atom coupled to a scalar field this has been done
previously in Refs. \cite{lambshift}.
We mentioned in section 3 that the shift can be
identified with the imaginary part of the $\rho_{eg}$ coefficient
(in the case of a two-level atom) in Eq. (\ref{rhoeg}), namely
\begin{equation}
\Delta E = \frac{-1}{\hbar}\mbox{Im } \big \{ K_{eg,ge}(
  -\omega_{eg}) + K_{ge,eg}^*(\omega_{eg}) \big \}
  = \frac{-|\vec{d}_{eg}|^2}{\hbar} \big \{ \mbox{ Im } K(-\omega_{eg})
    - \mbox{ Im } K(\omega_{eg}) \big \}\; .
\end{equation}
For the derivation of Im $K(\omega)$ we write it as
$[K(\omega) - K^*(\omega)]/(2i)$ and make use of Eqs. (\ref{gijende})
and (\ref{distrib}). This procedure results in
\begin{eqnarray}
  \mbox{Im } K(\omega)&=&\frac{\hbar a \omega^3}{12 \pi^2 \varepsilon_0 c^4}
  \int_0^\infty dt \cos(\omega t) \frac{{\cal P}}{\sinh(\Delta) \cosh^3
  (\Delta)} \Big \{ 1+ \frac{a^2}{\omega^2 c^2 \cosh^2(\Delta)} \Big \}
  \nonumber\\ & &
  + \frac{\hbar a^4}{96 \pi^2 \varepsilon_0 c^7} \int_0^\infty
  \frac{dt}{\cosh^4(\Delta)} \Big \{ 3 \sin(\omega t) \Big [ 2 \tanh^2 (\Delta
  ) - \frac{3}{\cosh^2(\Delta)} - \frac{8 \omega^2 c^2}{a^2}\Big ]
  \nonumber \\ & & \hspace{4cm}
  - \frac{22 \omega c}{a} \tanh(\Delta) \cos (\omega t) \Big \}
\end{eqnarray}
The first of the two integrals diverges for $t\rightarrow 0$, but the
divergence is the same as for the inertially moving atom. After the
use of the renormalization condition Im$K^{\mbox{\scriptsize inertial}}
(\omega)=0$ it is therefore well defined.

To see this more explicitely
we consider the case that the acceleration $a$ is smaller than $c
\omega_{eg}$. It is then a very good
approximation to expand the integrand to lowest order in $a$. We then find
\begin{equation}
  \mbox{Im } K(\omega) = \left ( 1+ \frac{a^2}{\omega^2 c^2}\right )
  \mbox{Im } K^{\mbox{\scriptsize inertial}}(\omega) - \frac{\hbar a^2
  \omega^2}{12 \pi^2 \varepsilon_0 c^5} \int_0^\infty \Big \{
  \frac{5 \omega t}{6} \cos (\omega t) + 3 \sin (\omega t)\Big \}
  +O(a^4)\; .
\end{equation}
Performing the integrations and applying the renormalization condition
we eventually arrive at
\begin{equation}
 \mbox{Im } K(\omega) = - \frac{13}{72\pi^2} \frac{\hbar \omega a^2}{
 \varepsilon_0 c^5} +O(a^4)\; .
 \label{imk}
\end{equation}
The corresponding {\em Lamb shift}
of the atoms due to the acceleration is given by
\begin{equation}
\Delta E = - \frac{13}{36 \pi^2} \frac{a^2 \omega_{eg} e^2
  |\langle e| \vec{x}| g \rangle |^2}{\varepsilon_0 c^5}
  +O(a^4)\; .
\label{lambsh} \end{equation}
It coincides structurally with the results of
Refs. \cite{lambshift} for the scalar coupling.
Although this result is interesting for theoretical reasons
it shows that the effect is far beyond the scope of present
experiments: Taking for $\hbar \omega_{eg}$ a value of 10 eV and
for the matrix element $|\langle e| \vec{x}| g \rangle |$ the
Bohr radius of $5\cdot 10^{-11}$m the Lamb shift in the Earth's
gravitational field ($a= 9.81 $ m/s$^2$) is of the order
of $10^{-55}$ eV.
%%%%%%%%%%%%%%%%%%%%%%%%%%%%%%%%%%%%%%%%%%%%%%%%%%%%%%%%%%%%%%%%
\section{Acceleration-induced multi-level effects}
We want to address three questions: As compared with the two-level atom, does
a multi-level atom show new effects caused by the acceleration? What is the
order of magnitude of these effects?
Are there states which show no spontaneous excitation at all?
To discuss this we study a three-level
atom because the answers can already be read off there.

{\em Three-level atom in $\Lambda$ configuration:} The
$\Lambda$ system consists of two ground states $|+ \rangle$ and
$|- \rangle$ with energies $E_+$ and $ E_-$ and an excited state $|e \rangle$
with energy $E_e$. The ground states are not directly coupled
so that only $d^{\underline{a}}_{+e}= d^{\underline{a}*}_{e+}$
and $d^{\underline{a}}_{-e}= d^{\underline{a}*}_{e-}$ are
nonvanishing.
The $\Lambda$ system is of interest because of the possibility
of Raman transitions (see, e.g., Ref. \cite{marzlin96}) between
the two ground states via the excited state under the influence of an
electromagnetic radiation field. These transitions
can be very effective even if the driving mechanism, for instance a laser,
is far off-resonant.

Turning to accelerated atoms in the Minkowski vacuum,
we display the master equation for $\rho_{++}$ and $\rho_{+-}$,
\begin{eqnarray}
\frac{d \rho_{++}}{dt} &=& \frac{1}{\hbar^2} \Big \{-2 \rho_{++}
  |\vec{d}_{+e}|^2 \mbox{ Re } K(\omega_{e+})
  + 2\rho_{ee} |\vec{d}_{+e}|^2 \mbox{ Re } K(-\omega_{e+})
  \nonumber \\ & & \hspace{0.5cm}-
  \rho_{-+} e^{i \omega_{+-}t}(\vec{d}_{+e}\cdot \vec{d}_{e-})
  K(\omega_{e-}) -
  \rho_{+-} e^{-i \omega_{+-}t} (\vec{d}_{+e}\cdot \vec{d}_{e-})^*
  K^*(\omega_{e-}) \Big \}
  \label{lambdasys} \\
\frac{d \rho_{+-}}{dt} &=& \frac{-1}{\hbar^2} \Big \{ \rho_{+-}
  \big [ |\vec{d}_{+e}|^2 \mbox{ Re } K(\omega_{e+}) +
         |\vec{d}_{-e}|^2 \mbox{ Re } K^*(\omega_{e-}) \big ]
  + e^{-i \omega_{+-}t} (\vec{d}_{+e}\cdot \vec{d}_{e-}) \times
  \nonumber \\ & & \hspace{1cm}
  \big [ \rho_{--} K(\omega_{e-}) + \rho_{++} K^*(\omega_{e+}) - \rho_{ee}
  \big ( K(-\omega_{e+}) + K^*(-\omega_{e-})\big ) \big ]
  \Big \} \label{lambdasyspm}
\end{eqnarray}
The corresponding equation for $\rho_{--}$ is gained if the + and - indices
in Eq. (\ref{lambdasys}) are interchanged. $\rho_{-+}$ is simply the
complex conjugate of $\rho_{+-}$, and the evolution of $\rho_{ee}$ is
described by $\dot{\rho}_{ee} = - \dot{\rho}_{++} - \dot{\rho}_{--}$.

According to Eq. (\ref{rhogg}) the first line
of Eq. (\ref{lambdasys}) agrees with the differential
equation of a two-level atom with levels $|+ \rangle$ and $| e \rangle$.
Because of Eqs. (\ref{rek}) and (\ref{imk}) and $\omega_{e-} >0$ the
second line in Eq. (\ref{lambdasys}) is a new acceleration-induced effect
typical for the multi-level situation.
It vanishes for $a=0$. The population $\rho_{++}$ of the $|+ \rangle$ state
is coupled to the coherences $\rho_{+-}$ between the two ground states.
The transition matrix elements $\vec{d}_{+e}$ and $\vec{d}_{-e}$ between
each of the lower states and the excited state appear so that we have
something similar to a
Raman transition between the two not directly coupled ground states.
On the other hand, the contribution of the two-point functions contained in
$K(\omega_{e-})$ are those of the two-level system with states $|e \rangle$
and $|- \rangle$. After renormalization of Im$K(\omega)$ as done above,
$K(\omega_{e-})$ is given by Eqs. (\ref{rek}) and (\ref{imk}). This has the
direct consequence that not only the first term in (\ref{lambdasys}) but
also the last two terms are of the order of the Unruh effect for two-level
atoms. Accordingly, although new effects appear, with regard to an experimental verification of the influence of acceleration the situation
is the same as in the two-level case. In this sense
the Unruh effect is not enhanced.

This reasoning holds also in the general multi-level case. Considering Eq.
(\ref{rhomaster}) one can see that the argument of any of the functions
$K_{ij,kl}(\omega)$ is always the frequency difference $\omega_{kl}$
of two states $| k \rangle$ and $| l \rangle$ which are coupled by a
dipole moment $d^{\underline{a}}_{kl}$.
Hence, the transition rate is always the same as the
one between the two states $| k\rangle$ and $| l\rangle$.
We have as an important result that
the magnitude of the acceleration-induced transition rates
for the multi-level atom are essentially the same as for
a two-level atom consisting of the states  $| k \rangle$ and $|l \rangle$.

Let us now turn to another observation.
Under special circumstances this multi-level effect has the interesting
consequence that a linear combination of the two ground states may remain
stable even in the presence of acceleration. For the $\Lambda$ system at hand
this happens if the dipole moments fulfill the condition
$|\vec{d}_{+e}|^2 = |\vec{d}_{-e}|^2 = |\vec{d}_{+e}\cdot \vec{d}_{e-}|$.
The {\em non-coupled state} then can be written as
\begin{equation}
|\psi_{nc} \rangle = \frac{1}{\sqrt{2}} \big \{ |+ \rangle - e^{-i \varphi}
  |- \rangle \big \}\; ,
\end{equation}
where the phase $\varphi$ is given by arg $(\vec{d}_{+e}\cdot \vec{d}_{e-})$.
It is not difficult to show that the corresponding density matrix
$\rho = |\psi_{nc} \rangle \langle \psi_{nc} |$ is time independent so
that there is no acceleration induced spontaneous excitation which makes
this state unstable. The Unruh effect appears only for the {\em coupled state}
\begin{equation}
|\psi_{c} \rangle = \frac{1}{\sqrt{2}} \big \{ |+ \rangle +e^{-i \varphi}
  |- \rangle \big \}\; ,
\end{equation}
for which spontaneous excitation is present.

The existence of coupled and non-coupled ground states is well known in the
field of atom optics and has led to an intense research in laser cooling of
atoms (see, e.g., Ref. \cite{ct92}) and dark optical lattices
(see, e.g., Ref. \cite{haensch}).
However, while in this field the coupling to laser
beams with given intensity and frequency is concerned, we are here dealing
with the coupling to the vacuum modes in the presence of acceleration which
acts similar to a thermal bath. We therefore conclude that although
the Unruh effect in general affects the stability of atomic levels there
is the possibility that the transition amplitudes of two different ground
states interfere destructively so that a combination of them may remain
stable.
\\[2mm]
%%%%%%%%%%%%%%%%%%%%%%%%%%%%%%%%%%%%%%%%%%%%%%%%%%%%%%%%%%%%%%%%
{\em Three-level atom in V configuration:}
Let us finally follow an other strategy which presents itself in connection
with multi-level atoms. Are there certain stable atomic configurations
where the stability depends crucially on the fact that the atom moves
inertially?
We study a possible candidate for such a situation. In the V configuration
of a three-level atom we have
one ground state $|g \rangle$ and two excited states
$|\pm \rangle$ for which we will assume $E_+=E_-$.
The excited states are not directly coupled so
that only $d^{\underline{a}}_{g+}=d^{\underline{a}*}_{+g}$
and $d^{\underline{a}}_{g-}=d^{\underline{a}*}_{-g}$
do not vanish. Because the two emssion amplitudes from each
excited level to the ground state can interfere destructively, the
total spontaneous emission rate was observed to be cancelled if
the two excited states are coupled to a fourth level by a coherent
driving field \cite{xia96}. For the sake of simplicity we will
consider here the unrealistic case that $d^{\underline{a}}_{-g}
=-d^{\underline{a}}_{+g}$ holds and for which theoretically
spontaneous emission cancellation occurs. Abbreviating
$K(\pm \omega_{+g})$ by $K_\pm$ the master equation
(\ref{rhomaster}) for the excited states can be found to be
\begin{eqnarray}
\frac{d \rho_{++}}{dt} &=& \frac{|\vec{d}_{g+}|^2}{\hbar^2} \Big \{
   2 \rho_{gg}\mbox{ Re }K_+
  -2 \rho_{++} \mbox{ Re } K_- +
  \rho_{-+} K_- +\rho_{+-} K^*_- \Big \} \\
\frac{d \rho_{--}}{dt} &=& \frac{|\vec{d}_{g+}|^2}{\hbar^2} \Big \{
  2 \rho_{gg} \mbox{ Re }K_+ -2 \rho_{--} \mbox{ Re }K_-
  +\rho_{+-} K_- +\rho_{-+} K_-^* \Big \}\\
\frac{d \rho_{+-}}{dt} &=& \frac{-|\vec{d}_{g+}|^2}{\hbar^2} \Big \{
  2 \rho_{+-}\mbox{ Re }K_-
  + 2 \rho_{gg}\mbox{ Re }K_+ -
  \rho_{--} K_- -
  \rho_{++} K_-^* \Big \}\; .
\end{eqnarray}
All we need to know about the other equations is that $d \rho_{gg}
/dt = - d \rho_{++}/dt - d \rho_{--}/dt$ and that the master equation
for the ground state coherences do not depend on $\rho_{\pm \pm}$
and $\rho_{\pm \mp}$. It is not difficult to see that these equations
have the stationary solution $\rho_{++}=\rho_{--}=\rho_{+-}=
\rho_{-+}=1/2$ corresponding to the pure excited state $|\psi \rangle
= \{ |+ \rangle + | - \rangle \}/\sqrt{2}$. Since this solution
is independent on the precise form of $K(\omega)$ we see
that the atomic center-of-mass motion has no influence on the
stability of this stable excited state although the individual
decay rates may change.
%%%%%%%%%%%%%%%%%%%%%%%%%%%%%%%%%%%%%%%%%%%%%%%%%%%%%%%%%%%%%%%%
\section{Discussion}
We have calulated the spontaneous transition rates
of multi-level atoms,  coupled to the electromagnetic field, which move
with constant acceleration through the Minkowski vacuum. Corrections to the
inertial rates become important only for huge accelerations. For multi-level
systems new acceleration induced transitions occur, but they do not alter
the magnitude of the usual Unruh effect. For a three-level atom in
$\Lambda$ configuration these transitions resemble Raman transitions, but in
this case they are spontaneous because no photons are present in the
Minkowski vacuum.

It is a common belief that the spontaneous processes induced by the
accelerated motion lead to a general instability of all the states of the
multi-level atom. We have shown that in contrast to this specific linear
superpositions of the two ground states of a $\Lambda$ system represent
a non-coupled pure state which is stable, i.e., shows no Unruh effect.

In addition, a three-level atom in V configuration has a state which is a
non-coupled  excited state if the atom is moving inertially.
This stability {\em is not changed} in the case of accelerated motion, so that
there is again no influence.

An influence of acceleration on the Lamb shift, although much too small to
be measured, is always there. We have obtained the respective expression for
the electromagnetic coupling after the usual renormalization.

%%%%%%%%%%%%%%%%%%%%%%%%%%%%%%%%%%%%%%%%%%%%%%%%%%%%%%%%%%%%%%%%

{\bf Acknowledgement}: We thank F. Burgbacher, M. Holzmann,
C. L\"ammerzahl
and R. M\"uller for discussions and the Optik Zentrum
Konstanz for financial support.
%%%%%%%%%%%%%%%%%%%%%%%%%%%%%%%%%%%%%%%%%%%%%%%%%%%%%%%%%%%%%%%%
%\newpage


\begin{thebibliography}{99}
\bibitem{jacobiproof} The proof goes as follows: Using the well known
	\cite{itzykson80}
	relation $\int d^3k f({\bf k},k^0 = k)/k  = 2 \int d^4k \theta (k^0)
	\delta (k\cdot k) f(k)$
	it is not difficult to show that $k\cdot k = p\cdot p$ and,
	because $e_{\underline{0}}^\mu$ points towards the future,
	$\theta (k^0)=\theta (p^0)$. It remains to show that $d^4 p=
	d^4 k |\mbox{ det } e_{\underline{\alpha}}^\mu| $ is equal to
	$d^4k$. This is true because $e_{\underline{\alpha}}^\mu \eta_{\mu
	\nu} e_{\underline{\beta}}^\nu = \eta_{\underline{\alpha}
	\underline{\beta}}$ from which one directly can deduce
	$|\mbox{ det } e_{\underline{\alpha}}^\mu|=1$.
\end{thebibliography}
\end{document}